\documentstyle[11pt,paspconf]{article}

\begin{document}

\title{CNO Abundances in Dwarf and Spiral Galaxies}
\author{B E J Pagel}
\affil{Astronomy Centre, CPES, University of Sussex, Brighton BN1 9QJ, UK}
\begin{abstract}
CNO abundances in galaxies bear on issues of galactic evolution as well 
as stellar evolution and nucleosynthesis. Knowledge about them in dwarf 
and spiral galaxies depends mainly on emission lines from H~{\sc ii} regions, 
with information from stars and supernova remnants available for 
galaxies in the Local Group. 

Oxygen abundances 
%(recently recalibrated both for the Sun and for high-metallicity H {\sc ii} 
%regions) 
can be related to both local and 
global properties of the parent galaxies %such as gas fraction, 
%surface brightness, galaxy luminosity,  
%rotation speed and morphological type, and 
In  H {\sc ii} regions they 
range from 1/30th solar (using the new calibration) to slightly above 
solar. C/O is a more or less smoothly increasing function of O/H, 
flat at low metallicities and rising above 12+log(O/H) = 8.0 with a 
45$^{\circ}$ 
slope. N/O behaves quite similarly in the traditional `primary followed by 
secondary' style, but with a substantial scatter at least around O-abundance 
8.0.      C/N ratios show no clear trend with metallicity. 
\end{abstract}


\section{Introduction}

CNO abundances test stellar yields ---
C and N from a wide range of stellar masses, and oxygen
mainly from massive stars above 10$M_{\odot}$.  As predicted by Andr\' e 
Maeder (1992) and discussed recently
by Henry et al.\ (2000), carbon yields increase with
metallicity somewhat like a traditional `secondary' element because of
stellar mass loss and
nitrogen perhaps even more steeply as a secondary product of the CNO cycle
on carbon, although there is also a `primary' component at low
metallicities. Oxygen yields decrease with metallicity, at least
to some extent, again because of stellar mass loss.

Apart from the implications for stellar evolution and nucleosynthesis,
oxygen is the standard element for defining `metallicity' in gas-rich
galaxies and so is used also as a test of galactic chemical evolution
%in relation to such questions as the influence of inflows or outflows,
%variable or constant yields associated with variable or constant IMFs
%between and across galaxies, structural 
%implications of abundance
%gradients, luminosity-metallicity relations etc., 
and as a standard against which to judge abundances
of other elements such as iron in stars and C and N in
H {\sc ii} regions.   

How well do we know CNO abundances in external galaxies? Most 
data come from H {\sc ii} regions 
and (in the Local Group) supernova remnants.
For  H {\sc ii} regions  of sufficient surface brightness and not too large
metallicity, the standard method is to deduce the electron temperature
in the O$^{++}$ zone from the  line ratio 4363/5007, use
models to find that in the O$^ +$ zone and add the resulting
abundances of O$^ +$ and O$^{++}$ relative to H$^ +$. This is  
a lower limit, because of $\delta t^ 2$  
(Peimbert 1967). Recombination lines are insensitive to temperature 
(Esteban et al.\ 2002), but they are weak and the interpretation not completely 
straightforward 
%because IR fine-structure lines --- also insensitive to 
%temperature --- seem to agree better with the optical nebular  lines 
(Tsamis et al.\ 2002). In most cases
the effect appears to be small, probably under 0.1 dex, but there is
another correction for oxygen locked on grains,
which again might approach 0.1 dex. 
% depending on the relative abundances
%of Mg, Si and Fe, the precise chemistry and the extent to which solid
%grains survive in the ionized regions.

When 4363 is too weak to measure, as is normally
the case when metallicity approaches or exceeds solar, 
uncertainties multiply. Searle (1971) had the fundamental insight that the 
spectral changes exhibited by giant H {\sc ii} regions across Scd galaxies 
(Aller 1942) were due to a radial abundance gradient, and thus 
inspired various `empirical' methods of abundance estimation 
based on strong nebular lines alone. The most widely used is the 
notorious $R_{23}$ method 
%based on [O~{\sc ii}] + [O~{\sc iii}]
(Pagel et al.\ 1979, 1980),
calibrated against a model for S5 in M 101 by
Shields \& Searle (1978).  Investigations since then have shown
that the oxygen in S5 was overestimated (Kinkel \& Rosa
1994), as it has been in some other regions (Castellanos et al.\  
2002), so that
our calibrations have overestimated oxygen
abundances near solar, and attempted recalibrations by McCall et al.\  
(1985), Evans \& Dopita (1985) and McGaugh (1991) have various 
weaknesses
of their own, quite apart from the inevitable non-uniqueness of the
relationship (cf.\ Pilyugin 2000). There is some promise in newly 
developed strong-line
indices like S$_{23}$ (D\'{\i}az \& Perez-Montero 2000), S$_{234}$ 
(Oey \& Shields  
2000)  and especially 
in refinements of the $R_{23}$ method by Pilyugin
(2000, 2001a) in which the relative strengths of [O  {\sc ii}] and [O  
{\sc iii}] lines are taken into account in an empirical manner, 
reducing the scatter.  Until more of these
refinements have been applied, however, abundances near and 
above solar have to be taken with a grain of salt.

After oxygen, the next step 
is usually to estimate the N/O ratio. 
A common assumption, supported by many models at least at low 
metallicities, is that nitrogen is ionized to the same degree as oxygen 
(e.g.\ Garnett 1990), so that N/O = N$^ +$/O$^ +$.  
%deduced from the line intensity ratio [N  {\sc ii}] 6584/[O  {\sc ii}]  
%3727. 
This can actually be checked from infra-red data when available,  
%giving fine-structure transitions of [N  {\sc iii}] and [O {\sc iii}], 
e.g.\ in 30 Doradus (Lester et al.\ 1987). 
Carbon lacks prominent emission lines in the visible or near IR, 
so one has to use UV observations, usually of the intercombination 
lines C {\sc iii}] 1909 and C {\sc ii}] 2326.
%  and there are some worries 
%about reddening and photo-ionization modelling, with the large sensitivity 
%of these high-energy collisionally excited lines to the electron 
%temperature.  

  CNO abundances are also now being measured in stars, mostly B stars 
and supergiants, in the nearest galaxies. 
%, using 10m-class telescopes 
%and sophisticated non-LTE model atmosphere analyses. 
As in the Milky Way, comparisons between stellar 
and nebular abundances provide a useful check. Here one has to be 
rather careful for two reasons. 
\begin{enumerate} 
\item Stellar abundances are often expressed 
differentially relative to the Sun, but the solar goalposts are shifting, 
at least in the cases of C and O (Holweger 2001; 
Allende Prieto et al.\ 2001, 2002) and this has caused some  
confusion.  
%One example of this is the old argument as to whether the 
%oxygen:iron ratio in Magellanic Cloud supergiants is lower than solar. 
%It once appeared that it was, but then it was shown that Galactic 
%supergiants show the same effect (Luck \& Lambert 1992, Venn 
%1995). From the more recent work it now seems that at least some of the 
%effect was due to solar oxygen being overestimated, a point that is also 
%relevant to comparisons between the Sun and the local ISM.\footnote{Recently 
%Smith et al.\ (2002) have found a large dispersion in Fe/O ratio among  
%the most iron-rich LMC giants with [Fe/H]$\simeq -0.3$.} 
% `The fault, dear Brutus, is not in our stars!'         
 
\item In some hot main-sequence stars and supergiants, especially if 
rotating, the atmospheres have 
undergone some CNO processing, so to that extent it is the 
differences between stars and H  {\sc ii} regions that are of interest; 
but oxygen is usually unaffected.     
\end{enumerate} 
 
\section{The Local Group} 

\begin{table}
\caption{{\large  CNO abundances in local group galaxies}}
\label{table1}
\begin{tabular}{|r|c|c|c|l|}
\hline
&&&&\\
Object                 & C &   N &   O & Reference \\ \hline
%{\small
\underline{Local Galaxy:} &&&&\\
Sun                   & 8.4  &(7.8?) & 8.7 & All. Pr.\ et al 01,02\\
 "                    & 8.6  & 7.9 &8.7 & Holweger 01 \\
Orion nebula           &8.5 & 7.8 & 8.7 & Esteban et al 98\\
diffuse ISM           &     &     &8.6 to 8.7 & Meyer et al 98\\
cepheids            &8.0/8.6 &&8.3/8.9&Luck et al 98\\  \hline
\underline{M 31:}$\;\;\;\;\;\;\;\;\;\;\;\;$&&&&\\
H {\sc ii} reg.    &  &7.0/8.2  & 8.5/9.2 & Dennef. \& Kunth 81\\
SNR                &  &7.4/8.0  & 8.3/8.7 & Blair et al 82     \\
4 AF supergi     & 8.3 & & 8.8 & Venn et al.\ 00\\ \hline
\underline{M 33:}$\;\;\;\;\;\;\;\;\;\;\;\;\;$&&&&\\
H {\sc ii} reg.    & & 7.9$-.16R$&9.0$-.12R$&V\'{\i}lchez et al 88\\
SNR                & & 7.8$-.12R$& 8.8$-.07R$& Blair et al 85\\
B,A supergi&  & & 9.0$-.16R$ & Monteverde et al\ 97 \\ \hline
\underline{LMC:}$\;\;\;\;\;\;\;\;\;\;\;\;\;$ &&&&\\
H  {\sc ii} reg., SNR & 7.9 & 6.9 & 8.4 & Garnett 99 \\
cepheids          & 7.7/8.3 & & 8.0/8.8 & Luck et al 98\\
{\small PS 34-16 (early B)} &    7.1 & 7.5 & 8.4 & Rolleston\\
{\small LH 104-24 ("")}$\;\;\;$ &   7.5 & 7.7 & 8.5 & et al 96 \\
{\small NGC 1818/D1}    &        7.8 & 7.4 & 8.5 & Korn et al 02\\
{\small N2004 (4 early B)} &  8.1 & 7.0 & 8.4 &$\;\;$   "      "\\
4 F supergi           & 8.1 &     &     & Russell \& Bessell 89 \\ \hline
\underline{SMC:}$\;\;\;\;\;\;\;\;\;\;\;\;\;$ &&&& \\
H  {\sc ii} reg., SNR & 7.5 & 6.6 & 8.1 & Kurt et al 98\\
cepheids &         7.4/7.8 && 8.0/8.3 & Luck et al 98\\
10 A supergi&$<7.3$/$<8.7$&6.8/7.7&8.1& Venn 99\\
3 F supergi            &7.7      &      &8.1&Spite et al 89\\
2 F supergi            &7.8    &     &&  Russell \& Bessell 89 \\ \hline
\underline{NGC 6822:}$\;\;\;\;\;$ &&&&\\
H  {\sc ii} reg. &      &       6.6  & 8.3& Pagel et al 80\\
"  "$\;\;$            &       &            & 8.4& Pilyugin 01c\\
2 A supergi &      &         & 8.4 & Venn et al 01\\  \hline
%}
\end{tabular}
\end{table}

A selection of recent data is given in Table 1.    
Several points emerge from this table. 
%The Magellanic Clouds are the nearest external galaxies and consequently 
%there is a wealth of data for H {\sc ii} regions and young stars. 
%About 10 years ago in Heidelberg I gave a review on abundances in the 
%Clouds.  At that time an interesting article had just 
%appeared in the ESO {\em Messenger} entitled `Trouble in the Magellanic 
%Clouds' and I commented on this by referring to an old BBC programme 
%called `So you think you've got troubles!', because of various discrepancies 
%in stars and the ISM in our own neighbourhood. The picture now looks a 
%lot better, largely because of improvements in solar and stellar 
%abundance analyses, while the picture from emission nebulae is virtually 
%unchanged. 
 
\begin{enumerate} 
\item There is excellent agreement between the oxygen abundances 
measured in stars and  H {\sc ii} regions, brought about in the case 
of the Sun and Orion by a combination of downward revision in the Sun 
and the use of recombination lines in Orion.  There is equally good 
agreement between gas and young stars in the Magellanic Clouds. 
This is an improvement on the situation 10 years ago when the systematic 
uncertainties were such that one had to resort to the expedient of 
comparing like with like --- B stars with B stars, supergiants with 
supergiants,  H {\sc ii} regions with  H {\sc ii} regions etc (Russell \& 
Dopita 1992; Pagel 1993). 
  
\item While some of the B stars and supergiants in both Clouds show 
evidence of CN processing, others do not and 
these show excellent agreement with  H {\sc ii} regions also in C and N. 

\item The two Clouds have similar C/O and N/O ratios, the latter being  
distinctly lower than in the Sun and Orion. We shall see later 
that there are other gas-rich dwarf galaxies in the same range of 
oxygen abundance showing a wide range in N/O.  

\end{enumerate} 

Relevant data are also accumulating for older objects in the LMC, 
at least for oxygen --- from giants in globular clusters (Hill et al.\ 
2000) and from planetary nebulae (Dopita et al.\ 1997) --- with quite 
good overall agreement. Oxygen, like iron, shows a fairly clean 
age-metallicity relation, unlike the case for the solar neighbourhood
(Hill et al.\ Fig 4).   

\section{Irregular and blue compact galaxies} 

Irregular and blue compact galaxies are gas-rich objects, mostly of 
low luminosity but having appreciable star formation rates leading 
to bright H {\sc ii} regions  that have been intensively studied in recent 
years, not least in the search for primordial helium. Some stellar data 
are now available for NGC 6822 giving good agreement with H  {\sc ii} 
regions (Table 1) and a slight hint of an abundance gradient. 

However, some points are still controversial. 
\begin{enumerate} 
\item Luminosity-metallicity (i.e. O-abundance) relation. 
First noted by Lequeux et al.\ (1979) in the form of a mass-metallicity 
relation, and discussed further by 
Skillman et al.\ (1989a) and Richer \& McCall (1995), 
the relationship for d Irr has been challenged (e.g.\   
Hidalgo-G\' amez \& Olofsson 1998;  Hunter \& Hoffman 1990), but 
there are uncertainties in 
measuring the the weak 4363 line and Pilyugin (2001c) using a more robust 
strong-line method finds it to hold pretty well.  
%from 7.2 at $M_ B =-11$ to 8.4 (i.e.\ half the revised solar) at $M_ B = -18$. 
Blue compact galaxies do not fit, but if metallicity 
is plotted against velocity dispersion --- a measure of the total 
mass --- the scatter is much reduced (Kobulnicky \& Zaritsky 1999). 
This relation is very important from the point of view of galactic 
chemical evolution and is suggestive of increasing loss of enriched 
gas from shallower potential wells, but there are also other 
reasons, notably the  gas fraction (cf.\ van Zee et al.\  
1997; Pilyugin \& Ferrini 2000; Larsen et al.\ 2001; Garnett 2002).   
Tidal dwarf galaxies lie above the usual relation 
(Duc \& Mirabel 1998); there may 
also be a modest upward shift for galaxies in a dense environment, 
but this is not certain (V\'{\i}lchez 1995). 
 
\item Uniformity of abundances. How uniform are abundances across a 
galaxy or even one giant H {\sc ii} 
region? Does the neutral gas have the same metallicity as the  H {\sc ii} 
regions or are they self-enriched (cf.\ Kunth \& Sargent 1986)? 
For oxygen, the evidence mostly favours uniformity; 
the most famous case in which it was challenged was that of I Zw 18 
%where initial estimates of the O-abundance from the resonance line 1302 
%in the neutral gas made it 
%much lower than the already low abundance measured in  H {\sc ii} 
(see Kunth et al.\ 1994;  
%), but radio measurements of the Doppler width of 
% H {\sc i} showed that the lines are very saturated and there does not 
%actually seem to be any difference --- at least in the central regions (
Pettini \& Lipman 1995; van Zee et al.\ 1998c; Levshakov et al.\ 2001). 
Thuan et al.\  
(1997) found a very low upper limit to oxygen abundance in 
the H {\sc i} gas of the BCG SBS 0335 $-$052, but until the Doppler width 
is directly measured as in I Zw 18,  the significance of this limit remains 
in doubt.  

\item For nitrogen, the existence of intrinsic scatter is indisputable 
(e.g.\ Kobulnicky \& Skillman 1996, 1998; Kobulnicky et al.\ 1997), but 
the meaning 
of this scatter is unclear. In NGC 5253 and II Zw 40, large variations 
occur in N/H over small spatial intervals even within what looks like 
one giant H~{\sc ii} region, unaccompanied by noticeable variations in any 
other element (Walsh \& Roy 1989, 1993), whereas in others like NGC 4214 
there is an anti-correlation 
between N/O and O/H such as might be expected to occur if shortly after a 
burst of star formation oxygen were enhanced but nitrogen 
had not yet had time to catch up (Garnett 1990; 
Marconi, Matteucci \& Tosi 1994; Pilyugin 1993, 1999; Larsen et al.\  
2001). However, according to Izotov \& Thuan
(1999), below an oxygen abundance of about 7.7,  
%(one tenth of the newly revised solar), 
all the scatter that appeared in previous diagrams 
of H  {\sc ii} galaxies is spurious; N/O is remarkably constant and free 
from scatter at 1/40, compared to solar 1/7.  That figure also appears to 
be a well-defined lower limiting plateau  for all known  H {\sc ii} 
regions and possibly
representative of all irregular and BCG's not affected by whatever is 
causing the scatter around an oxygen abundance of 8.0, since it still applies 
to NGC 6822 at an oxygen abundance of 8.3 (cf Skillman et al.\ 1989b).   
  
The location of the scatter raises a problem for 
the point of view conventionally put forward for explaining it, 
based on a time delay for especially primary nitrogen (e.g.\ Edmunds \& 
Pagel 1978),  
since that theory requires the scatter to be greatest at the lowest 
metallicities, just the opposite of what is seen, and so IT argue that 
the primary nitrogen comes from massive stars and is produced in lockstep 
with oxygen, the scatter at higher metallicities coming then from 
secondary nitrogen, as was 
suggested by Pagel et al.\ (1986), although the association 
 with strong WR features that we suggested then has not been supported by 
later work.  Bursts, time lags and IMS  
could still be involved, however, if the lowest-metallicity galaxies 
are exceptionally slow evolvers rather than just very young, so that 
the effects of all past bursts have had time to be evened out, and 
there are some indications of scatter in N/O among the DLA clouds.  It could 
then be the regions with an oxygen abundance around 8.0, e.g. SMC and NGC 
4214, that show the greatest influence of bursts --- involving both 
primary and secondary nitrogen --- compared to still more metal-rich 
regions in spiral galaxies. I think the   
issue of how much massive stars contribute to primary nitrogen is still open  
(cf.\ Meynet \& Maeder 2002ab). 

%\begin{figure*}[htbp]
%\vspace{8cm} 
%\special{psfile=kob.ps angle=0 hoffset=-70 voffset=-140 vscale=60 hscale=80} 
%\caption{Trends of C/O and C/N with oxygen abundance, adapted from 
%Kobulnicky et al.\ (1997).} 
%\end{figure*} 
    
\item Carbon abundances show a plateau followed by a rising trend as a 
function of metallicity, with C/O increasing 
from about 1/5 in I Zw 18 to about 1/3 in the LMC  
(Garnett et al.\ 1997; Izotov \& Thuan 1999); see Fig 1. There is some 
dispute as to the smoothness of the trend,  
% Garnett et al.\ found 
%a modest amount of scatter at the lowest metallicities, which is 
%disputed by Izotov \& Thuan, the issue being the precise electron 
%temperature, 
but the differences for objects 
in common never 
exceed 0.1 dex. 
The rise in C/O with O/H probably  results from increasing carbon 
and decreasing oxygen yields from massive stars (Maeder 1992; Henry, Edmunds 
\& K\" oppen 2000).  
\end{enumerate}

\section{Spiral galaxies} 

\subsection{Oxygen}  
%Two complicating factors affect the discussion of oxygen abundances 
%in spiral galaxies. One is the existence of radial abundance gradients, 
%sometimes amounting to nearly an order of magnitude over the Holmberg 
%radius, and the other is the effect of environment, e.g.\ on Virgo cluster 
%galaxies compared to field galaxies.  In addition, the abundance in central 
%regions, estimated from the R$_{23}$ method, has probably been 
%overestimated and not many galaxies have been studied using the more 
%modern strong-line methods like Pilyugin's P-method. A few stellar data 
%exist for the nearest spirals and the agreement with H {\sc ii} regions 
%and supernova remnants as a function of position is quite good; see Table 1.   

Oxygen abundance gradients vary quite widely (e.g.\ Zaritsky et al.\  
1994; van Zee et al.\ 1998b): in late-type spirals like 
M 33 and M 101 (see Torres-Peimbert et al.\ 1989; Pilyugin 2001b), 
the variation is from more or less 
solar near the centre to about an order of magnitude lower at the isophotal 
radius, while in early-type spirals high abundances extend to greater 
radial distances. Also at the other end of the Hubble sequence, Sd and Sdm 
spirals and low surface brightness galaxies (de Blok \& van der Hulst 
1998) have shallower gradients again, with a lower abundance overall. 
McCall (1982), Edmunds \& Pagel (1984) and Vila-Costas \&
Edmunds (1992) studied correlations with local properties at the 
appropriate galactocentric distance, i.e.\ the surface brightness and 
the gas fraction, while  Zaritsky et al.\  stress the 
importance of global galaxy properties: the luminosity and the rotation 
speed (representing mass).  
%taking a representative oxygen abundance at 
%0.4 of the de Vaucouleurs isophotal radius, they find a 
%metallicity--luminosity relation similar to that for ellipticals. 
Probably the global properties plus any accidental interactions are 
the main driver, leading to the formation of exponential 
disks in which the correlations with local properties then arise.    
 
O-abundance 
correlates with local surface brightness and gas fraction, with some 
segregation between morphological types, and the mean abundance increases 
with the parent galaxy's luminosity or mass as indicated by the rotational 
velocity.   A further variable is the environment, e.g.\ in inner parts of 
the Virgo cluster there are gas-stripped galaxies with higher abundances 
at a fixed fraction of the isophotal radius than normal galaxies of the 
same type, but such cases can also be found among field galaxies (Pilyugin 
et al.\ 2001). 
%, and references therein). 

\subsection{Nitrogen} 

N/O ratios in spiral galaxy H {\sc ii} regions start distinctly above the 
Magellanic Cloud plateau value; e.g.\ at O=8.4, outer parts of spirals 
typically have log (N/O)=$-1.2$ compared to the LMC value of $-1.5$  
(Thurston et al.\  
1996; van Zee et al.\ 1998a; Pilyugin et al.\ 2001. )
There is some intrinsic scatter with early spirals having a slightly larger 
N/O at given O/H than late spirals (Pilyugin et al.\ 2002), but 
with the new calibrations 
the Sun and Orion fit in perfectly. We can see secondary nitrogen production 
taking over from primary at an oxygen abundance of about 8.5. 
 
\subsection{Carbon} 

Garnett et al.\ (1999) have used HST to measure carbon abundances in 
selected H {\sc ii} regions of M 101 and NGC 2403 and compared the 
results with irregular galaxies, Galactic H {\sc ii} regions and the 
Sun. Disregarding the scatter, which may be spurious, C/O behaves 
remarkably like N/O and C/N is nearly constant throughout the abundance 
range.
% (see Fig 1). 

\section{Putting it all together} 

A comprehensive discussion of CNO in irregular and spiral galaxies has 
been given by Henry et al.\ (HEK 2000).  They collected 
data from both stars and nebulae, so 
that there is a great mixture of ages there, but within the uncertainties 
the stars and the nebulae agree pretty well. Consequently, the nebulae 
convey the same message as the stars about carbon and oxygen, already 
discussed previously by Prantzos et al.\ (1994) and by 
Gustafsson et al.\ (1999). Inter\-mediate-mass stars are inadequate for the 
task of making carbon and one needs large metallicity-dependent yields 
from massive stars, as originally proposed by our birthday boy in 1992.  
Since massive stars are involved in both cases, 
the C/O ratio is largely a function of the stellar yields and insensitive 
to some galactic phenomena like the star formation rate. 

\begin{figure*}[htbp]
\vspace{9.cm} 
\includegraphics{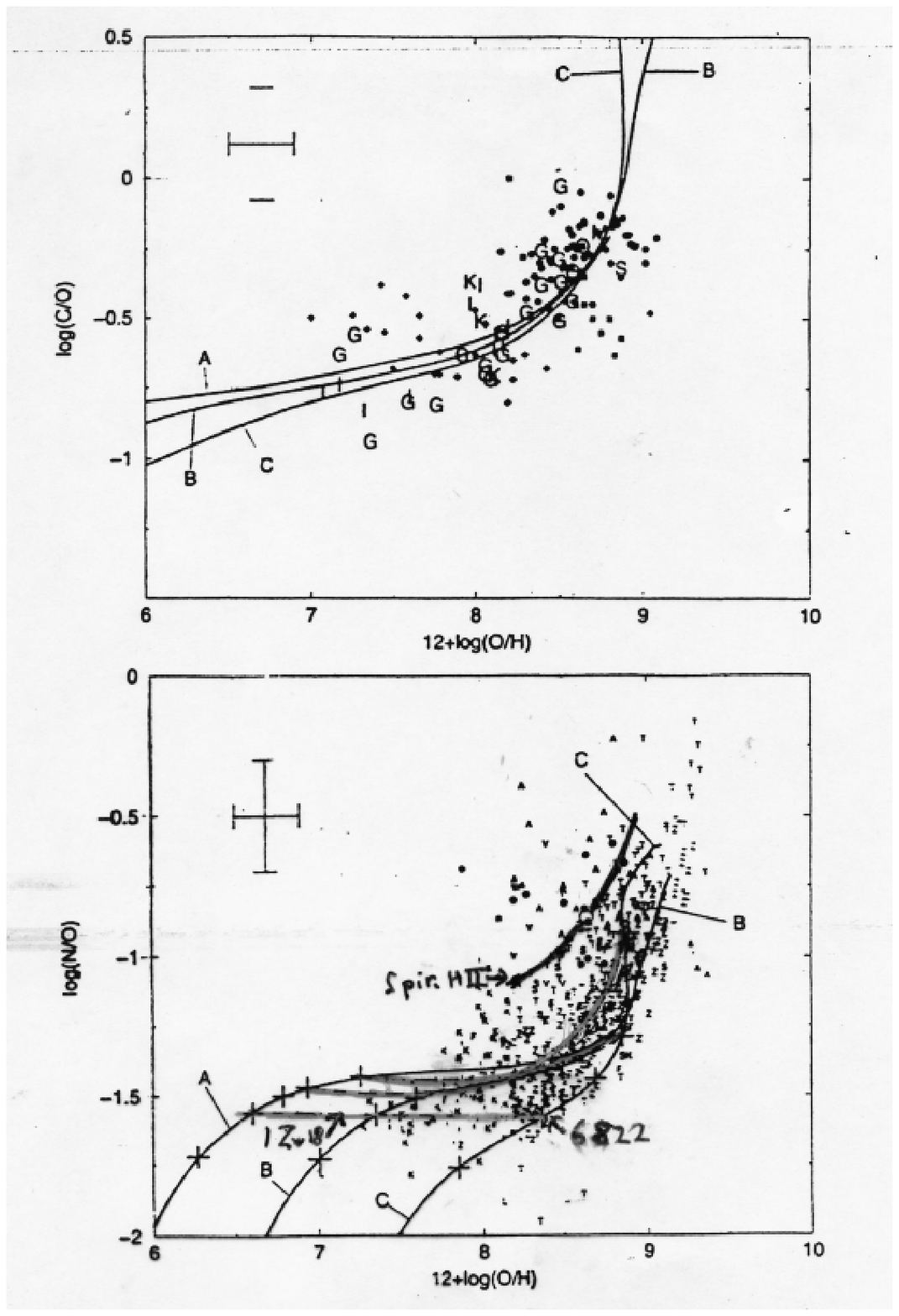} 
\caption{Fig 3b of HEK (2000) with isochrones (grey 
lines) and the 
results of Pilyugin et al.\ (2001) overplotted.} 
\end{figure*} 
    
The case of nitrogen is more complicated because of time delays  of up to 
$2\times 10^ 8$ years associated with the evolution of IMS,  
which produce enough primary nitrogen to account for the plateau at low 
metallicities (van den 
Hoek \& Groenewegen 1997; Meynet \& Maeder 2002ab).  The secondary component 
comes from the high-mass stars, again as predicted by Maeder (1992), 
but this time there is more sensitivity to galactic parameters such as 
the star formation rate (see Fig 1). Looking at the GCE model 
computations by HEK, 
one might expect the stars to lie along evolutionary tracks, but the 
H {\sc ii} regions could equally be along isochrones, or some mixture 
of the two. In fact the 
objects with log (N/O) $\simeq-1.6$ (from I Zw 18 to NGC 6822) lie pretty 
much along HEK's 0.5 Gyr isochrone (the exact age would be a function 
of the adopted star formation law), although HEK themselves preferred to 
consider them as lying along an evolutionary track. The isochrone 
interpretation is essentially the one suggested by Edmunds \& Pagel 
(1978). The 2 Gyr isochrone 
has a shape not very different from the Pilyugin et al.\ observational 
track for spirals, but is a factor of 2 or so too low, so it may be that 
one has to invoke inflow or some other effect not included in their models, 
and probably higher nitrogen yields from low-mass stars.

%\begin{references}
%\reference Kinkel, \& Rosa M. 1988, AA, 89, 427  
%\reference Headroom, M. 1988, \apj, 278, 356
%\end{references}


\begin{thebibliography} 
%{\small 
\bibitem{} Allende Prieto, C., Lambert, D.L. \& Asplund, M. 2001, 
ApJL, 556, L63  
\bibitem{} Allende Prieto, C., Lambert, D.L. \& Asplund, M. 2002, 
astro-ph/0206089 
\bibitem{} Aller, L.H. 1942, ApJ, 95, 52 
\bibitem{} Blair, W.P., Kirshner, R.P. \& Chevalier, R.A. 1982, ApJ, 254, 50 
\bibitem{} Blair, W.P., Kirshner, R.P. \& Chevalier, R.A. 1985, ApJ, 289, 582
%\bibitem{} Campbell, A., Terlevich, R.J. \& Melnick, J., 1986, MNRAS, 223, 811  
\bibitem{} Castellanos, M., D\'{\i}az, A. \& Terlevich, E. 2002, MNRAS, 
329, 315  
%\bibitem{} Consid\' ere, S., Coziol, R., Contini, T. \& Davoust, E. 2000, 
%AA, 356, 89 (N/O in starbursting barred spirals) 
\bibitem{} de Blok, W.J.G.\ \& van der Hulst, J.M. 1998, AA, 335, 421 
%\bibitem{} Denicol\' o, G., Terlevich, R. \& Terlevich, E. 2002, MNRAS, 330, 69 
\bibitem{} Dennefeld, M. \& Kunth, D. 1981, AJ, 86, 989  
\bibitem{} D\'{\i}az, A.I. \& Perez-Montero, E. 2000, MNRAS, 312, 130   
\bibitem{} Dopita, M.A., Vassiliadis, E., Wood, P.R. et al.\ 1997, 
ApJ, 474, 188 
\bibitem{} Duc, P.-A. \& Mirabel, I.F. 1998, AA, 333, 813 
\bibitem{} Edmunds, M.G. \& Pagel, B.E.J. 1978, MNRAS, 185, 77  
\bibitem{} Edmunds, M.G. \& Pagel, B.E.J. 1984, MNRAS, 211, 507 
\bibitem{}  Esteban, C., Peimbert, M., Torres-Peimbert, S. \& Escalante,
V. 1998, MNRAS, 295, 401  
\bibitem{}  Esteban, C., Peimbert, M., Torres-Peimbert, S. \& Rodriguez, M. 
2002, ApJ, in press, astro-ph/0208313   
\bibitem{} Evans, I.N. \& Dopita, M.A. 1985, ApJS, 58, 125   
\bibitem{} Garnett, D.R. 1990, ApJ, 363, 142 
\bibitem{} Garnett, D.R. 1999, in Y.-H. Chu, N.B. Suntzeff, J.E. Hesser \& 
D.A. Bohlender (eds), {\em New Views of the Magellanic Clouds}, IAU Symp. 
Series Vol. 190, Kluwer, Dordrecht, p. 266    
\bibitem{} Garnett, D.R. 2002, ApJ, astro-ph/0209012  
\bibitem{} Garnett, D.R., Shields, G.A., Skillman, E.D., Sagan, S.P. \& 
Dufour, R.J. 1997, ApJ, 489, 63  
\bibitem{} Garnett, D.R. et al. 1999, ApJ, 513, 168  
\bibitem{} Garnett, D.R., Skillman, E.D., Dufour, R.J. \& Shields, G. 
1997, ApJ, 481, 174  
\bibitem{} Gustafsson, B., Karlsson, T., Olsson, E., Edvardsson, B. \& 
Ryde, N. 1999, AA, 342, 426 
\bibitem{} Henry, R.B.C., Edmunds, M.G. \& K\" oppen, J. 2000, ApJ, 541, 660  
\bibitem{} Hidalgo-G\' amez, A.M. \& Olofsson, K. 1998, AA, 334, 45 
\bibitem{} Hill, V., Fran\c cois, P., Primas, F. \& Spite, F. 2000, 
AA, 364, L19  
\bibitem{} Holweger, H. 2001, in R.F. Wimmer-Schweingruber (ed), 
{\em Solar and Galactic Composition}, AIP Conf. Proc., astro-ph/0107426   
\bibitem{} Hunter, D.A. \& Hoffman, L. 1999, AJ, 117, 2789 
\bibitem{} Izotov, Y. \& Thuan, T.X. 1999, ApJ, 511, 639  
\bibitem{} Kinkel, U., \& Rosa, M. 1994, AA, 282, 37  
\bibitem{} Kobulnicky, H.A. \& Skillman, E.D. 1996, ApJ, 471, 211
\bibitem{}  Kobulnicky, H.A. \& Skillman, E.D. 1998, ApJ, 497, 601  
\bibitem{}  Kobulnicky, H.A., Skillman, E.D., Roy, J.-R., Walsh, J.R. 
\& Rosa, M.R. 1997, ApJ, 477, 679  
\bibitem{}  Kobulnicky, H.A. \& Zaritsky, D. 1999, ApJ, 511, 118 
\bibitem{} Korn, A.J., Keller, S.C., Kaufer, A. et al.\ 2002, astro-ph/ 
0201453
\bibitem{} Kunth, D., Lequeux, J., Sargent, W.L.W. \& Viallefond, F. 
1994, AA, 282, 709  
\bibitem{} Kunth, D. \& Sargent, W.L.W. 1986, ApJ, 300, 496  
\bibitem{} Kurt, C.M., Dufour, R.J., Garnett, D.R. et al.\ 1998, preprint   
\bibitem{} Larsen, T.I., Sommer-Larsen, J. \& Pagel, B.E.J. 2001, 
MNRAS, 323, 555 
\bibitem{} Lequeux, J., Peimbert, M., Rayo, J.F., Serrano, A. \&
Torres-Peimbert, S. 1979, AA, 80, 155  
\bibitem{} Lester, D.F., Dinerstein, H.L., Werner, M.W. et al.\ 1987, 
ApJ, 320, 573  
\bibitem{}  Levshakov, S.A., Kegel, W.H. \& Agafonova, I.I. 2001, AA, 373, 836 
%\bibitem{} Luck, R.E. \& Lambert, D.L. 1992, ApJS, 74, 303 
\bibitem{} Luck, R.E., Moffett, T.J., Barnes, T.G. \& Gieren, W.P. 1998, 
AJ, 115, 605  
\bibitem{} McCall, M.L. 1982, Thesis, Univ. Texas, Austin  
\bibitem{} McCall, M.L.,  Rybski, P.M. \& Shields, G.A. 1985, ApJS, 57, 1   
\bibitem{} McGaugh, S.S. 1991, ApJ, 380, 140 
\bibitem{} Maeder, A. 1992, AA, 264, 105  
\bibitem{} Marconi, A., Matteucci, F. \& Tosi, M. 1994, MNRAS, 270, 35  
\bibitem{} Meyer, D.M., Jura, M. \& Cardelli, J.A. 1998, ApJ, 493, 222 
\bibitem{} Meynet, G. \& Maeder, A. 2002a, AA, 381, L25; astro-ph/0111187 
\bibitem{} Meynet, G. \& Maeder, A. 2002b, astro-ph/0205370  
\bibitem{} Monteverde, M.I., Herrero, A., Lennon, D.J. \& Kudritzki, R.-P. 
1997, ApJ, 474, L107 
\bibitem{} Oey, M.S. \& Shields, J.C. 2000, ApJ, 539, 687      
\bibitem{} Pagel, B.E.J. 1993, in B. Baschek, G. Klare, \& J. Lequeux (eds.), 
{\em New Aspects of Magellanic Cloud Research}, Springer-Verlag, p. 330.  
\bibitem{} Pagel, B.E.J., Edmunds, M.G., Blackwell, D.E., Chun, M.S. 
\& Smith, G. 1979, MNRAS, 189, 95 
\bibitem{} Pagel, B.E.J., Edmunds, M.G. \& Smith, G. 1980, MNRAS, 193, 219 
\bibitem{} Pagel, B.E.J., Terlevich, R.J. \& Melnick, J. 1986, PASP, 98, 1005  
\bibitem{} Peimbert, M. 1967, Ap. J., 150, 825  
\bibitem{} Pettini, M. \& Lipman, K. 1995, AA, 297, L63    
\bibitem{} Pilyugin, L. S. 1993, AA, 277, 42  
\bibitem{} Pilyugin, L. S. 1999, AA, 346, 428  
\bibitem{} Pilyugin, L. 2000, AA, 362, 325 
\bibitem{} Pilyugin, L. S. 2001a, AA, 369, 594 
\bibitem{} Pilyugin, L. S. 2001b, AA, astro-ph/0105103  
\bibitem{} Pilyugin, L. S. 2001c,  AA, 374, 412, astro-ph/0105360  
\bibitem{} Pilyugin, L. \& Ferrini, F. 2000, AA, 358, 72  
\bibitem{} Pilyugin, L. S., Moll\' a, M., Ferrini, F. \& V\'{\i}lchez, J.M. 
 2001, astro-ph/0112128  
\bibitem{} Pilyugin, L.S., Thuan, T.X. \& Vilchez, J.M. 2002, AA, 
astro-ph/0210225 
\bibitem{} Prantzos, N., Vangioni-Flam, E. \& Chauveau, S. 1994, 
AA, 285, 132 
\bibitem{} Richer, M.G. \& McCall, M.L. 1995, ApJ, 445, 642  
\bibitem{} Rolleston, W.R.J., Brown, P.J.F., Dufton, P.L. \& Howarth, I.D. 
1996, AA, 315, 95  
\bibitem{} Russell, S.C. \& Bessell, M.S. 1989, ApJS, 70, 865 
\bibitem{} Russell, S.C. \& Dopita, M.S. 1992, ApJ, 384, 508 
\bibitem{}Searle, L. 1971, ApJ, 168, 327    
%\bibitem{}Searle, L., Wilkinson, A. \& Bagnuolo, W.G. 1980, ApJ, 239, 803 
\bibitem{} Shields, G. \& Searle, L. 1978, ApJ, 222, 821  
\bibitem{} Skillman, E.D., Kennicutt, R. \& Hodge, P.W. 1989a, ApJ, 347, 875    
\bibitem{} Skillman, E.D., Terlevich, R. \& Melnick, J. 1989b, MNRAS, 240, 563 
%\bibitem{} Smith, V.V., Hinkle, K.H., Cunha, K.\ et al.\ 2002, 
%astro-ph 0208417  
\bibitem{} Spite, M., Barbuy, B. \& Spite, F. 1989, AA, 222, 35  
\bibitem{} Thuan, T.X., Izotov, Y.I. \& Lipovetsky, V.A. 1997, ApJ, 
477, 661  
\bibitem{} Thurston, T.R.,  Edmunds, M.G. \& Henry, R.B.C. 1996, MNRAS, 
283, 990   
\bibitem{} Torres-Peimbert, S., Peimbert, M. \& Fierro, J. 1989, 
ApJ, 345, 186 
\bibitem{} Tsamis, Y.G., Barlow, M.J., Liu, X.-W., Danziger, I.J. \& Storey, 
P.J. 2002, MNRAS, in press, astro-ph/0209534 
\bibitem{} van den Hoek, L.B. \& Groenevegen, M.A.T. 1997, AAS, 
123, 305  
\bibitem{} van Zee, L., Haynes, M.P. \& Salzer, J.J. 1997, AJ, 114, 2497  
\bibitem{} van Zee, L.,  Salzer, J.J. \& Haynes, M.P. 1998a, ApJ, 497, L1 
\bibitem{} van Zee, L.,  Salzer, J.J.,  Haynes, M.P., O'Donoghue, A.A, 
\& Balonek, T.J. 1998b, AJ, 116, 2805  
\bibitem{} van Zee, L., Westpfahl, D., Haynes, M.P. \& Salzer, J.J. 1998c, 
AJ, 115, 1000    
%\bibitem{} Venn, K.A. 1995, ApJ, 449, 839 
\bibitem{} Venn, K.A., 1999, ApJ, 518, 405  
\bibitem{} Venn, K.A., McCarthy, J.K. Lennon, D.J., Przybilla, N. 
 Kudritzki, R.P. \&   Lemke, M. 2000, ApJ, 541, 610  
\bibitem{} Venn, K.A., Lennon, D.J., Kaufer, A., McCarthy, J.K., 
Przybilla, N., Kudritzki, R.P,   Lemke, M,  Skillman, E.D. \&  
Smartt, S.J. 2001, ApJ, 547, 765  
\bibitem{} V\'{\i}lchez, J.M. 1995, AJ, 110, 1090 
\bibitem{} V\'{\i}lchez, J.M., Pagel, B.E.J., D\'{\i}az, A.I., 
Terlevich, E. \& Edmunds, M.G. 1988, MNRAS, 235, 633  
\bibitem{} Vila-Costas, M.B. \& Edmunds, M.G. 1992, MNRAS, 259, 121  
\bibitem{} Walsh, J.R. \& Roy, J.-R. 1989, MNRAS, 239, 297 
\bibitem{} Walsh, J.R. \& Roy, J.-R. 1993, MNRAS, 262, 27 
\bibitem{} Zaritsky, D., Kennicutt, R.C. Jr \& Huchra, J.P. 1994, 
ApJ, 420, 87 
%} 
\end{thebibliography}
\end{document}